\documentclass[a4paper,10pt]{article}
\pdfoutput=1
\usepackage{jcappub}
\usepackage[utf8]{inputenc}
\usepackage[T1]{fontenc}
\usepackage{todonotes}
\usepackage{ulem}
\usepackage{amsmath}
\usepackage[separate-uncertainty = true]{siunitx}
\DeclareSIUnit\gauss{G}

\usepackage{booktabs}

\usepackage{lmodern}

\DeclareRobustCommand{\permil}{%
  \ifmmode
    \text{\textperthousand}%
  \else
    \textperthousand
  \fi}
\DeclareSIUnit\gauss{G}

\usepackage{lineno,hyperref}
\usepackage{subcaption}

\usepackage{soul}

\title{Observation of inclined EeV air showers with the radio detector of the Pierre Auger Observatory}

\author[75]{A.~Aab,}
\author[67]{P.~Abreu,}
\author[49,48]{M.~Aglietta,}
\author[18]{I.F.M.~Albuquerque,}
\author[12]{J.M.~Albury,}
\author[1]{I.~Allekotte,}
\author[8,11]{A.~Almela,}
\author[63]{J.~Alvarez Castillo,}
\author[74]{J.~Alvarez-Mu\~niz,}
\author[40,42]{G.A.~Anastasi,}
% ...

\affiliation[1]{Centro At\'omico Bariloche and Instituto Balseiro (CNEA-UNCuyo-CONICET), San Carlos de Bariloche, Argentina}
% ...
\affiliation[8]{Instituto de Tecnolog\'\i{}as en Detecci\'on y Astropart\'\i{}culas (CNEA, CONICET, UNSAM), Buenos Aires, Argentina}
% ...
\affiliation[11]{Universidad Tecnol\'ogica Nacional -- Facultad Regional Buenos Aires, Buenos Aires, Argentina}
\affiliation[12]{University of Adelaide, Adelaide, S.A., Australia}
% ...
\affiliation[18]{Universidade de S\~ao Paulo, Instituto de F\'\i{}sica, S\~ao Paulo, SP, Brazil}
% ...
\affiliation[40]{Gran Sasso Science Institute, L'Aquila, Italy}
% ...
\affiliation[42]{INFN Laboratori Nazionali del Gran Sasso, Assergi (L'Aquila), Italy}
% ...
\affiliation[48]{INFN, Sezione di Torino, Torino, Italy}
\affiliation[49]{Osservatorio Astrofisico di Torino (INAF), Torino, Italy}
% ...
\affiliation[63]{Universidad Nacional Aut\'onoma de M\'exico, M\'exico, D.F., M\'exico}
% ...
\affiliation[67]{Laborat\'orio de Instrumenta\c{c}\~ao e F\'\i{}sica Experimental de Part\'\i{}culas -- LIP and Instituto Superior T\'ecnico -- IST, Universidade de Lisboa -- UL, Lisboa, Portugal}
% ...
\affiliation[74]{Instituto Galego de F\'\i{}sica de Altas Enerx\'\i{}as (I.G.F.A.E.), Universidad de Santiago de Compostela, Santiago de Compostela, Spain}
\affiliation[75]{IMAPP, Radboud University Nijmegen, Nijmegen, The Netherlands}
% ...
\affiliation[c]{also at Vrije Universiteit Brussel, Brussels, Belgium}
% ...

\emailAdd{auger\_spokespersons@fnal.gov}

\abstract{With the Auger Engineering Radio Array (AERA) of the Pierre Auger Observatory, we have observed the radio emission from 561 extensive air showers with zenith angles between 60$^\circ$ and 84$^\circ$.
In contrast to air showers with more vertical incidence, these inclined air showers illuminate large ground areas of several km$^2$ with radio signals detectable in the 30 to 80\,MHz band.
A comparison of the measured radio-signal amplitudes with Monte Carlo simulations of a subset of 50 events for which we reconstruct the energy using the Auger surface detector shows agreement within the uncertainties of the current analysis.
As expected for forward-beamed radio emission undergoing no significant absorption or scattering in the atmosphere, the area illuminated by radio signals grows with the zenith angle of the air shower.
Inclined air showers with EeV energies are thus measurable with sparse radio-antenna arrays with grid sizes of a km or more.
This is particularly attractive as radio detection provides direct access to the energy in the electromagnetic cascade of an air shower, which in case of inclined air showers is not accessible by arrays of particle detectors on the ground.
}

\keywords{cosmic rays, inclined air showers, radio detection, Pierre Auger Observatory, AERA}
\arxivnumber{1806.05386}

\begin{document}

\maketitle

\flushbottom

\section{Introduction}
Over the last 10 years, radio detection of extensive air showers has matured from small prototype setups to full-fledged applications as part of established cosmic-ray observatories~\citep{HuegePLREP}.
The technique relies on the measurement of coherent radio emission dominantly arising from time-varying transverse currents induced by the Earth's magnetic field, with secondary radiation arising from the time variation of the negative charge excess present in air showers~\citep{HuegePLREP}.
While radio measurements in the very-high-frequency band, typically from 30 to 80\,MHz, have reached competitive precision in the reconstruction of the important air-shower parameters arrival direction, energy, and depth of shower maximum~\citep{HuegePLREP,SchroederReview} in the energy range from ${\sim}10^{17}$\,eV to ${\sim}10^{18}$\,eV, also an intrinsic limitation has become apparent:
The radio emission from air showers with zenith angles up to ${\sim}60^\circ$ illuminates areas with diameters of only a few hundred meters, i.e., for coincident detection of such air showers with at least three radio antennas, an antenna grid with a spacing of order 200 to 300\,m is needed.
As the radio-emission footprint does not grow significantly with the energy of the primary particle, the antenna grid spacing would need to be equally dense for detection of near-vertical air showers with energies well beyond an EeV, which is obviously problematic as the low cosmic-ray flux at these energies requires instrumentation of very large areas.

There have been long-standing predictions that inclined air showers with zenith angles of more than $60^{\circ}$ should illuminate much larger areas and thus be more favorable for detection with sparse antenna arrays~\citep{GoussetRavelRoy2004,HuegeFalcke2005b}.
The reason for this expectation is that for inclined showers, the shower, and with it the source of the radio emission, is significantly further away from the ground than for near-vertical showers, while there is no relevant absorption or scattering of the radio signals in the atmosphere.
The strongly forward-beamed radio emission thus illuminates a significantly larger area, even after projection effects are corrected for.
As the total radiation energy in the radio signal is then distributed over a larger area, the signal strengths are weaker than for non-inclined air showers, which leads to an increase of the energy threshold for detection in the presence of noise.
(The background in the 30 to 80\,MHz band is dominated by radio emission from the Galaxy~\citep{HuegePLREP,SchroederReview}, and as such irreducible.)
Recent simulation studies with full Monte-Carlo simulations have confirmed these earlier findings and further illustrated the potential of radio meausrements of inclined air showers~\citep{HuegeUHECR2014}.

Radio signals from inclined air showers have previously been detected in the frequency range from 40 to 80\,MHz with the small-area LOPES experiment~\citep{PetrovicApelAsch2006}.
However, no quantitative analysis of the signal distributions was possible with the limited data available.
The ANITA experiment has measured several inclined cosmic-ray events at frequencies between 200 and 1200\,MHz after reflection of the radio signals off the Antarctic ice~\cite{HooverNamGorham2010}, and has even derived a flux at an energy of 2.9\,EeV~\citep{ANITAEnergy}.
The measurements at only one antenna location, however, did not allow validation of the radio-emission simulations that the analysis relied upon.
Lately, the ARIANNA experiment has reported the measurement of radio signals from cosmic rays in the frequency range of 100 to 500\,MHz, with one event measured at a zenith angle of approximately 75$^{\circ}$~\citep{Barwick:2016mxm}.

In this work we demonstrate that the Auger Engineering Radio Array (AERA)~\citep{SchulzIcrc2015}, in spite of not having been optimized for the detection of inclined air showers, has observed inclined air showers at EeV energies with significant statistics.
This is possible because, indeed, the radio emission of inclined air showers is detectable over areas of several km$^2$, providing the potential to measure such air showers even with antenna arrays on grid spacing of a km or more.
In addition, we show that the absolute radio-signal amplitudes measured at up to 76 antenna locations per air shower are in agreement with state-of-the-art Monte-Carlo simulations, and that we observe a signature of the Cherenkov time compression arising from the non-unity refractive index of the atmosphere.
We first describe our experimental setup and data analysis, then present our results, and finally discuss the potential of radio detection of inclined air showers in future applications.

\section{Experimental setup and data analysis} \label{sec:setup}

The Pierre Auger Observatory is a multi-hybrid detector for the measurement of ultra-high-energy cosmic rays~\citep{AugerNIM2014}, situated in the province of Mendoza, Argentina.
Its baseline detectors comprise a surface-detector (SD) array of 1660 water-Cherenkov particle detectors covering an area of 3000\,km$^2$, and a fluorescence detector (FD) consisting of 27 optical telescopes monitoring the atmosphere above the surface detector for ultraviolet fluorescence light. The fluorescence detector is used in particular to set the energy scale of the measurements, and allows direct observation of the depth of maximum of individual air showers. The Auger Engineering Radio Array is situated in the north-western part of the observatory.

AERA has been set up in stages.
In the period between June 2013 and March 2015 it consisted of a total of 124 antenna stations covering an area of 6\,km$^2$ employing a graded layout with grid spacings between 150\,m and 375\,m.
In the analysis presented here, those 76 out of the 124 antenna stations which are capable of receiving an external trigger from the surface detector array were used.
The area instrumented by these antennas amounts to approximately 3.5\,km$^2$.
Two different types of antennas are being used in AERA: logarithmic-periodic dipole antennas~\citep{AERAAntennaPaper2012,AERALPDA2017} and butterfly antennas~\citep{AERAAntennaPaper2012}.
Both measure the radio signal in the 30 to 80\,MHz band and have sensitivity at elevations below 30$^\circ$, but were never optimized for the detection of inclined air showers.
In particular, these antennas only measure the projection of the three-dimensional electric-field vector onto the horizontal plane, i.e., they do not measure the vertical field component, which can contain a significant fraction of the signal in case of inclined air showers~\citep{ApelArteagaBaehren2012a}.
Also, the systematic uncertainties in the absolute antenna calibration at low elevations have yet to be studied in detail and are currently larger than for near-vertical geometries.
Another limitation of the current setup with respect to inclined air showers lies in the selection of radio-readout triggers from the triggers provided by the surface detector.
Only events for which the closest (ground) distance between a surface detector station with a local trigger and an AERA station is smaller than 5\,km currently trigger the readout of the radio-antenna array.
This condition is applied to improve the purity of the acquired data sample, but is not tailored for horizontal air showers.
(The non-availability of the shower zenith angle at this stage of event processing currently makes it difficult to optimize this trigger for inclined air showers.)
Overall, there is thus still significant potential for improving the radio detection of inclined air showers with AERA.

\begin{figure}[htb]
  \centering
  \includegraphics[width=0.7\textwidth]{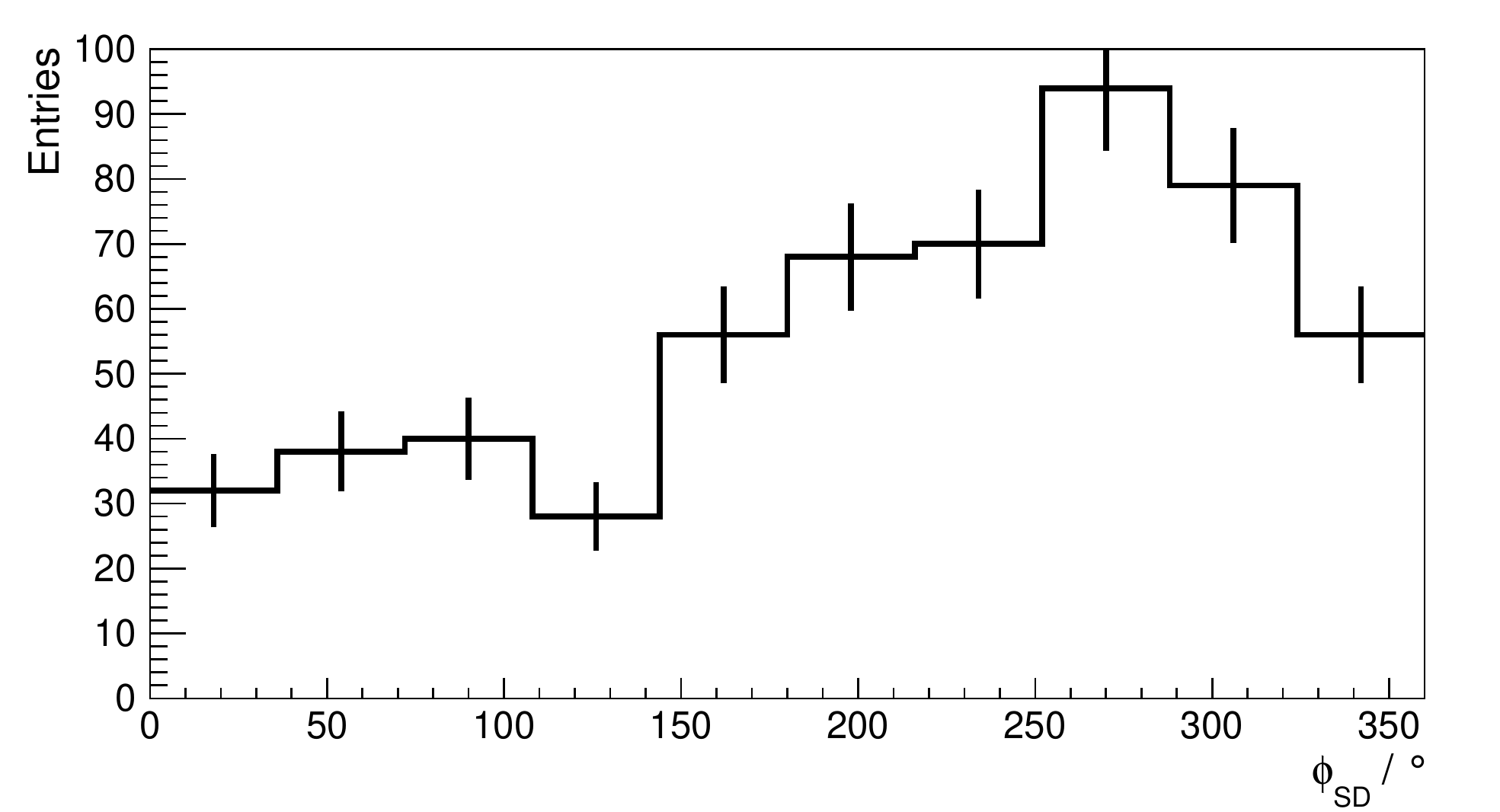}
  \includegraphics[width=0.7\textwidth]{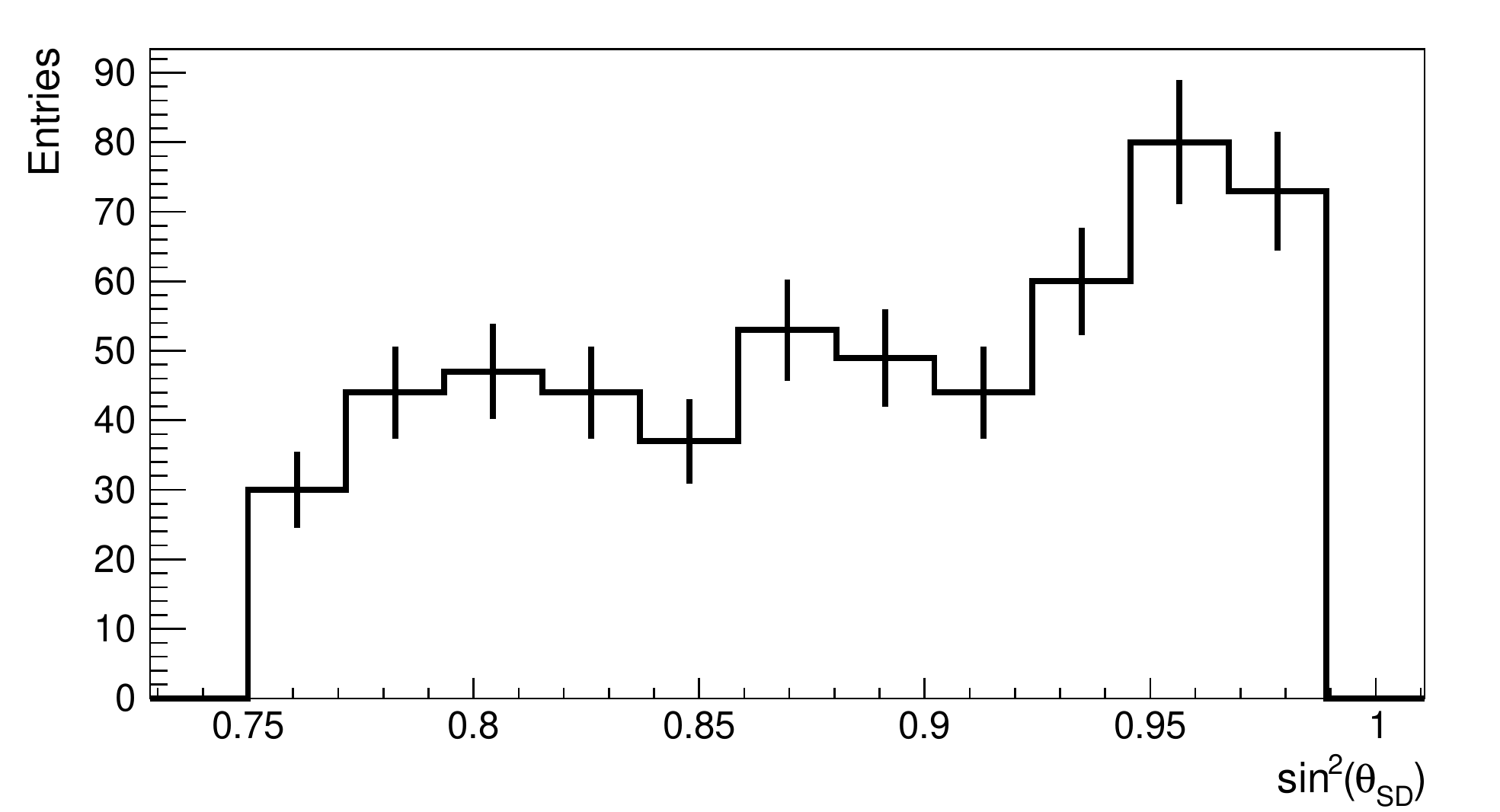}
  \caption{Distributions of the arrival directions of the 561 extensive air showers selected in this analysis as determined with the Auger surface detector. Top: Distribution of the azimuth angles, 0$^\circ$ indicating arrival from the east and counting counter-clockwise. Bottom: Distribution of the $\sin^2$ of the zenith angles. For each bin, Poissonian errors are shown.}
  \label{fig:angulardistribution}
\end{figure}

We analyzed a data set recorded between 26 June 2013 and 28 February 2015 employing a hybrid analysis of the surface-detector and radio-detector data with the Offline analysis framework of the Pierre Auger Observatory~\citep{ArgiroOffline2007}.
For the particle component of the air shower, we apply the standard reconstruction for inclined showers with zenith angles greater than 60$^\circ$ measured with the 1500\,m surface-detector array~\citep{AugerSDHAS}.
While the maximum zenith angle of this reconstruction is normally limited to 80$^\circ$ to ensure bias-free reconstruction parameters, we here extend the analysis to zenith angles up to 84$^\circ$, resulting in a minor degradation in reconstruction performance not relevant for the analysis presented here.
We refrain from analyzing events with zenith angles larger than $84^{\circ}$ at this time, as this would require additional studies and optimizations of the surface-detector reconstruction.
The radio analysis uses the event geometry determined with the surface-detector reconstruction as a starting point and requires coincident detection with a signal-to-noise ratio\footnote{Square of the maximum electric-field amplitude after projection onto the horizontal plane divided by square of the RMS of the background electric-field amplitudes.} of 10 or more in at least three radio-antenna stations with radio-pulse arrival times in approximate agreement with the event geometry~\citep{KambeitzARENA2016,KambeitzThesis2016}.
The combined surface-detector and radio analysis results in a total of 561 events with zenith angles between $60^{\circ}$ and $84^{\circ}$.

The distribution of the arrival directions as determined with the surface detector is shown in Fig.~\ref{fig:angulardistribution}.
As expected, the azimuthal distribution shows a clear north-south asymmetry: more air showers are observed coming from the south, where the large angle to the Earth's magnetic field leads to strong geomagnetic radio emission.
The distribution with respect to the $\sin^2$ of the zenith angle $\theta$ is not flat, as would be expected for a planar detector observing an isotropic flux with full efficiency.
The increase towards larger values of $\sin^2\theta$ indicates that the detection efficiency for the coincident observation of air showers with the Auger surface detector and AERA increases with zenith angle.
The mean zenith angle of the selected events amounts to 71$^{\circ}$.

Reconstructing the arrival directions of the air showers with a plane-wave fit to the arrival times of the radio pulses yields an average agreement to within 1.4$^{\circ}$ with the directions reconstructed from the surface-detector measurements.

\begin{figure}[htb]
  \centering
  \includegraphics[width=0.7\textwidth]{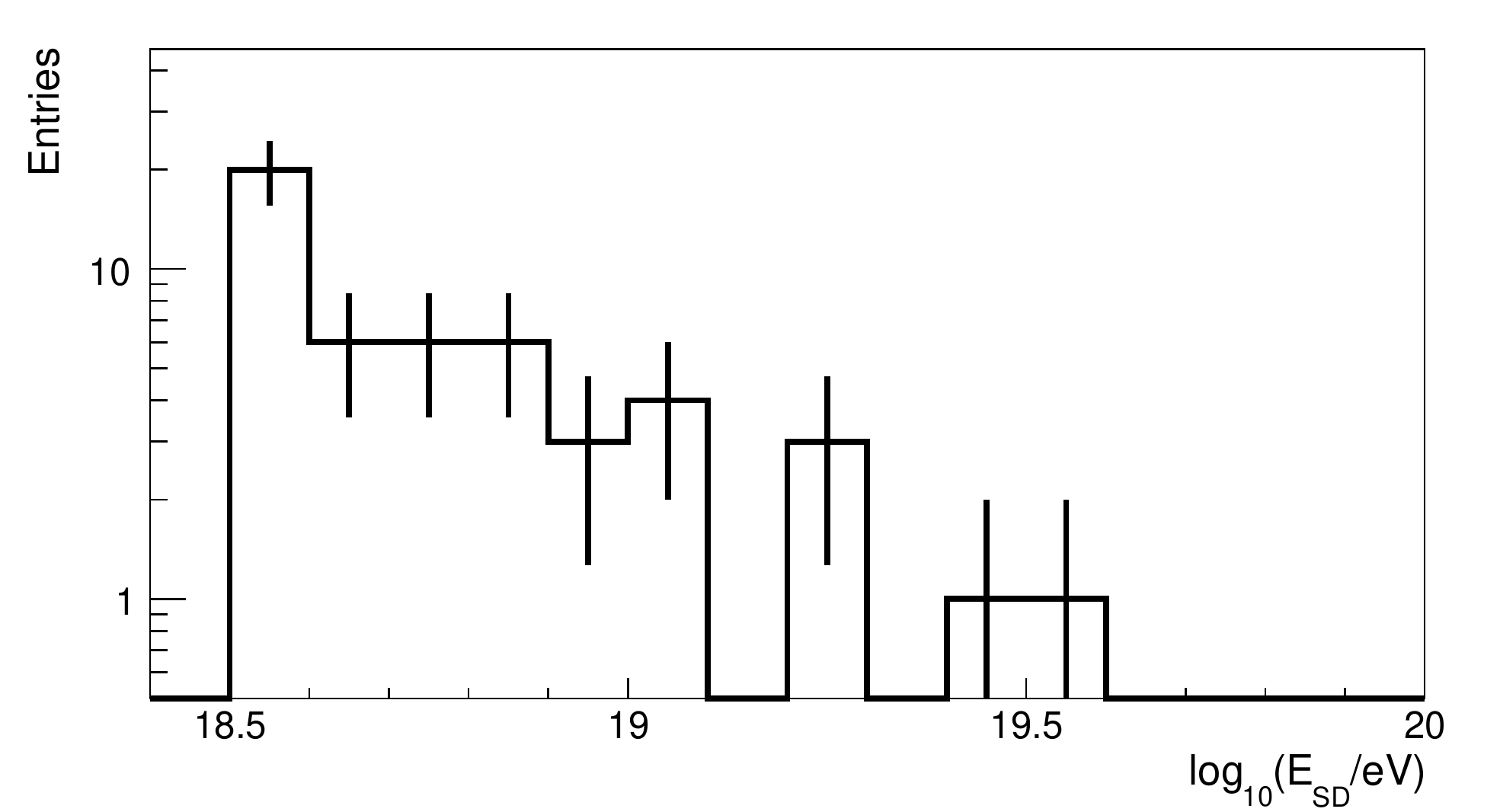}
  \caption{Distribution of the energy reconstructed with the surface detector for the subset of 50 events passing quality cuts for the energy determination with the Auger surface detector (see text). For each bin Poissonian errors are shown.}
  \label{fig:energydistribution}
\end{figure}

For a subset of the 561 events, the surface-detector reconstruction for inclined air showers~\citep{AugerSDHAS} allows us to determine the cosmic-ray energy after the calibration with the fluorescence detector. This is normally done for showers with zenith angles between $60^{\circ}$ and $80^{\circ}$, signals measured in a complete hexagon~\citep{Abraham:2010zz} of surface detector stations around the station with the largest energy deposit (thus in particular excluding events for which the impact point is not contained inside the area of the surface detector), and a minimum reconstructed energy of $10^{18.6}$\,eV.
This energy determination is bias-free and achieves a resolution of 19.3\%~\citep{AugerSDHAS}.
As discussed above, we here extended the zenith-angle range to include showers with zenith angles up to 84$^\circ$. At the same time, we lowered the energy threshold to $10^{18.5}$\,eV.
The requirements for detectors with measured signals remained unchanged.
Under these conditions, the bias of the sample remains negligible, and the energy resolution remains better than 25\%~\citep{DembinskiThesis2009}.
The loss of performance is not relevant for the analysis presented here.
The reconstruction with these selection criteria leads to a total of 50 events.
The energy distribution of this subset of events is shown in Fig.~\ref{fig:energydistribution}, illustrating the reach up to energies beyond $10^{19}$\,eV.

For each of these 50 events, one full Monte-Carlo simulation has been performed with CoREAS~\citep{HuegeARENA2012a} as part of CORSIKA~v7.56~\citep{HeckKnappCapdevielle1998} and using the interaction models QGSJET-II.04~\citep{Ostapchenko:2010vb} and UrQMD~1.3.1 \citep{Bleicher:1999xi}.
The event geometries and cosmic-ray energies for the simulations were taken from the surface-detector reconstruction, and the primary particle type was set to protons, yielding a good coverage of shower-to-shower fluctuations.
We did not simulate heavier primary particles because the effect of the primary mass on the predicted average radio amplitudes is well below 10\% (as given by the relative difference in the energy fraction in the electromagnetic cascade, cf.\ reference~\citep{Pierog:2015epa}).
Thus, the dominating uncertainties in the comparison between data and simulations are the energy-scale and calibration uncertainties (see section~\ref{sec:simcomparison}).
The simulation results have been propagated through a detailed radio-detector simulation~\citep{AbreuAgliettaAhn2011}, including the addition of radio noise measured at the time of the individual events, and have then undergone the same radio-data analysis as the measured data.

%%%%%%%%%%%%%%%%%%%%%%%%%%%%%%%%%%%%%%%%%%%%%%%%%%%

\section{Results}

In the following, we present results related to the size of the radio-emission footprint, the agreement between data and simulations, and the signature of a Cherenkov time-compression of the radio signals due to the non-unity refractive index of the atmosphere.

\begin{figure}[tbp]
  \centering
  \includegraphics[width=0.49\textwidth]{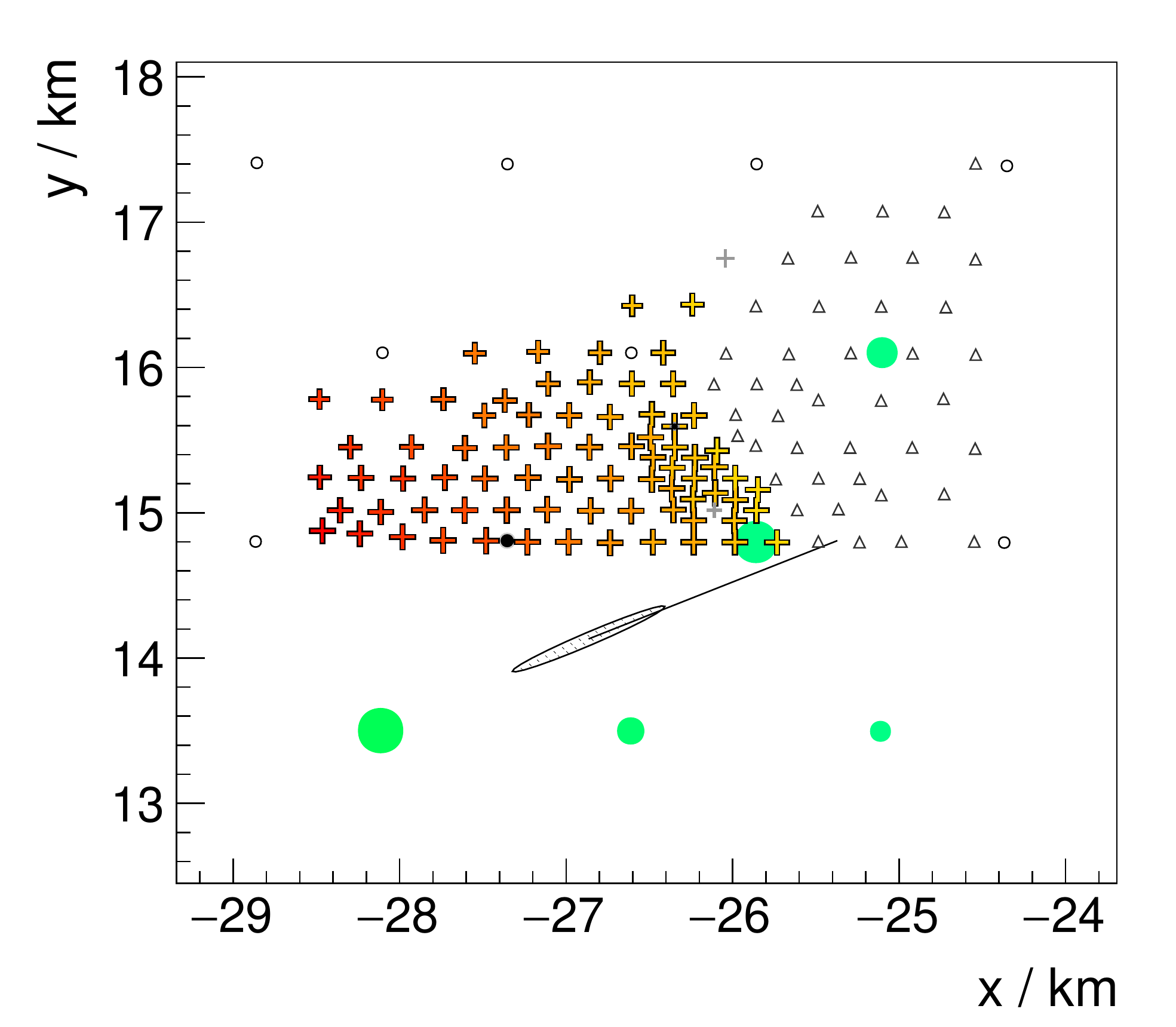}
  \includegraphics[width=0.49\textwidth]{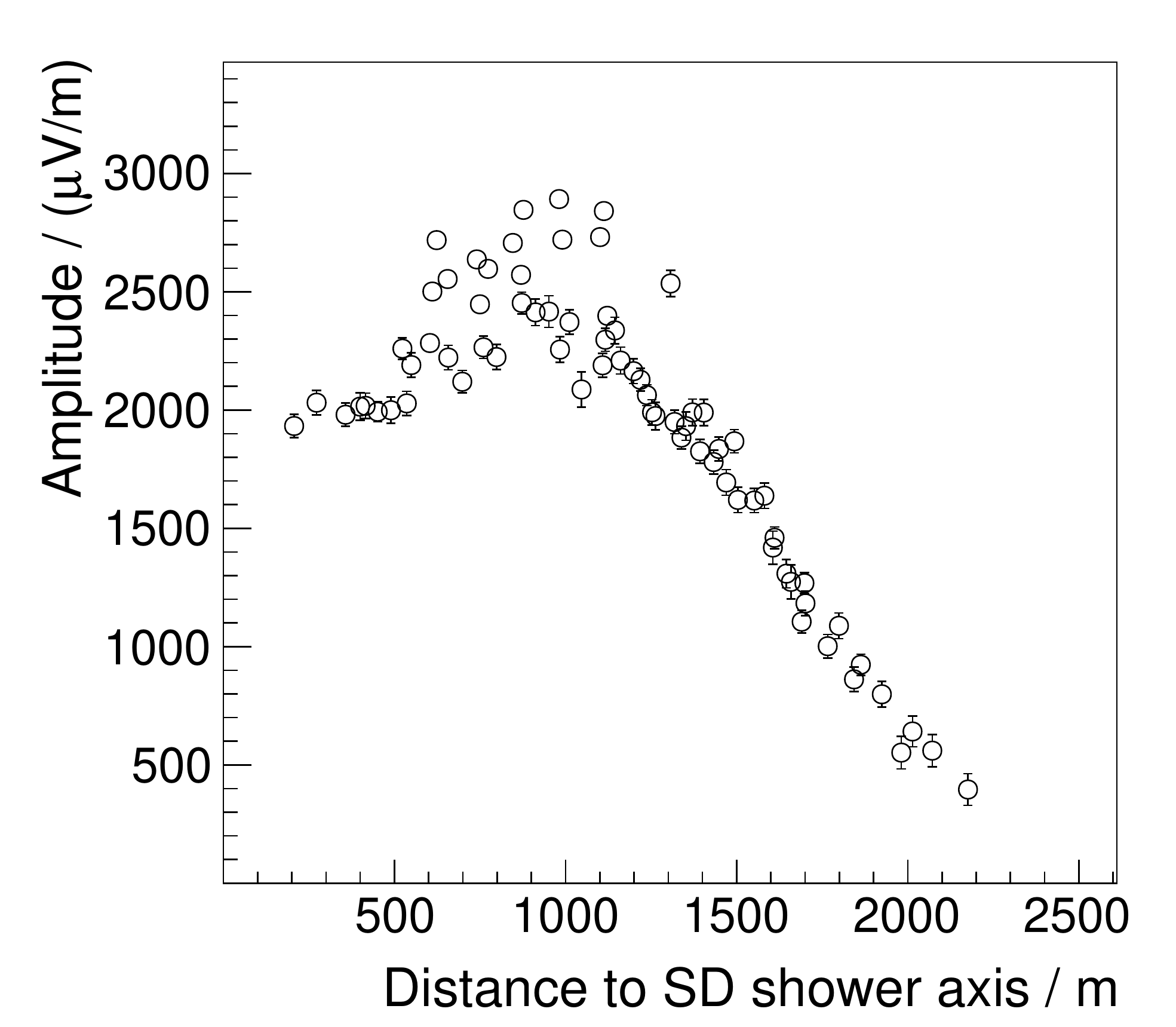}
  \caption{Visualisation of one event arriving at a zenith angle of $\theta=82.8^{\circ}$ from 24.3$^{\circ}$ north of east with an energy of $2{\times}10^{19}$\,eV. Left: View of the radio antennas with detected signal indicated by crosses color-coded from early (yellow) to late (red) arrival. Particle detectors with a signal are indicated by green circles, their size indicating energy deposit. Further particle detectors with a signal are present outside the shown area. The particle detector station marked in black had a temporary malfunction. Sub-threshold radio-detector stations are marked with grey crosses, radio stations not read out (in particular those not externally triggered) are denoted by triangles. The impact point as reconstructed with the surface-detector stations is indicated by the one-sigma error ellipse. The line indicates the projection of the shower axis onto the ground. Right: Amplitude distribution of the horizontal component of the electric field vector as a function of the distance from the shower axis measured in the plane normal to the shower axis. 74 radio-detector stations have a signal-to-noise ratio of 10 or higher.
  }
  \label{fig:exampleevent}
\end{figure}

\begin{figure}[htbp]
  \centering
  \includegraphics[width=0.7\textwidth]{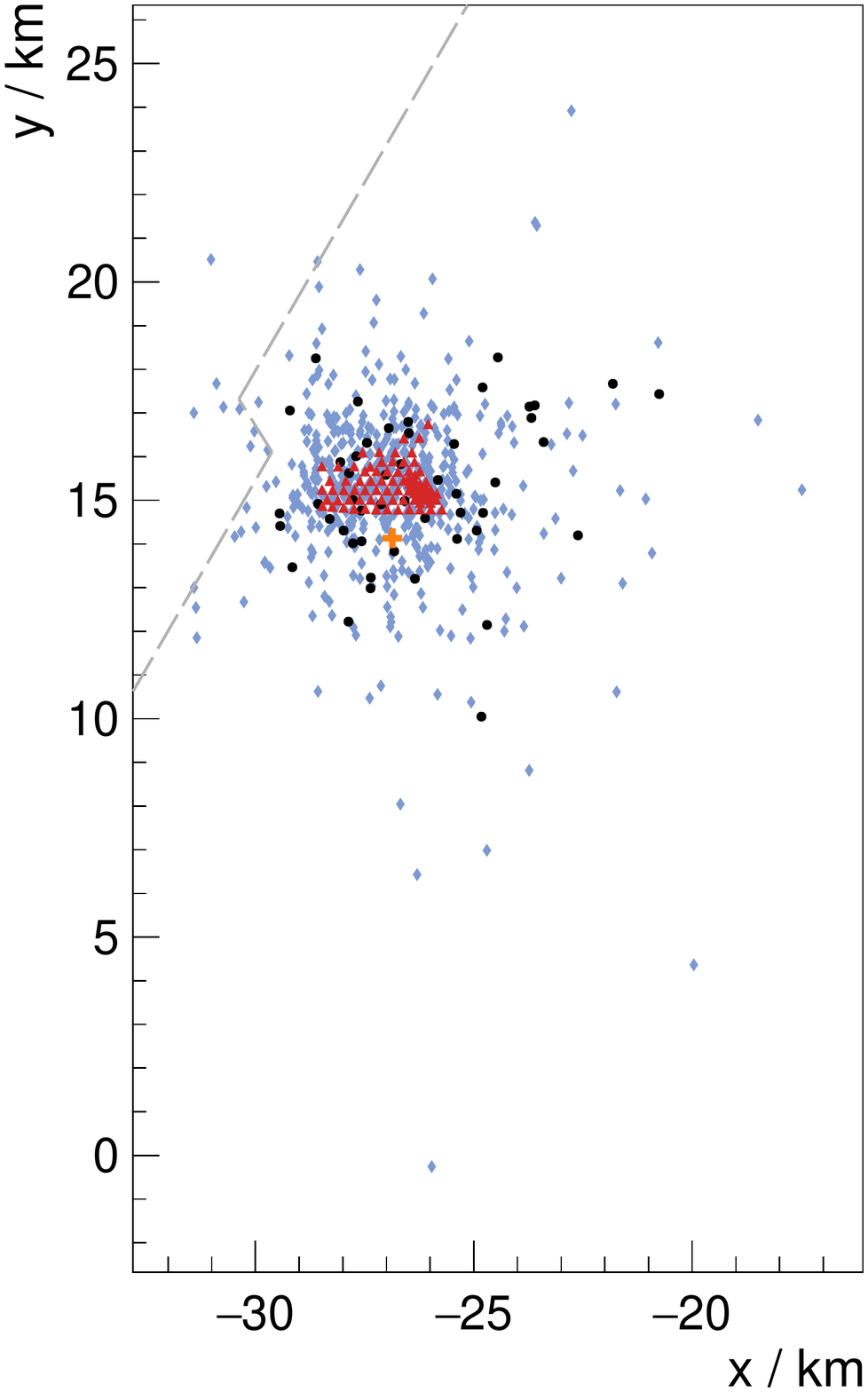}
  \caption{Distribution of the impact points of the air-shower axes reconstructed with the Auger surface detector for the 561 selected events, relative to the positions of the AERA antennas (red triangles). The 50 events that pass the quality cuts for energy reconstruction are marked by black dots, the remaining 511 events by blue diamonds. The example event presented in Fig.~\ref{fig:exampleevent} is marked with an orange cross. Fewer events with impact points in the west (left border of the plot) are present as this region is close to the edge of the Auger surface detector array, indicated by the dashed line.}
  \label{fig:coredistribution}
\end{figure}
\subsection{Size of the area illuminated by radio signals}

An illustration of the large size of the area illuminated by radio signals in inclined air showers is shown for one high-energy, high zenith-angle example event in Fig.~\ref{fig:exampleevent}.
While near-vertical air showers are typically detected in 3-5 AERA antennas~\citep{AERAEnergyPRD}, here a total of 74 antenna stations have a signal-to-noise ratio above 10 in the horizontal component of the electric field.
(The two events with the highest station multiplicity in the selection have a signal above threshold in all 76 antenna stations.)
The signal amplitudes measured in this event rise up to a distance of approximately 1000~m from the shower axis\footnote{Distance from the air-shower axis as reconstructed with the surface detector, measured in the plane perpendicular to the axis.} and then fall off to large distances, yet signals have been detected above the Galactic background up to axis distances of 2200\,m.
The illuminated area in the plane perpendicular to the shower axis for this event amounts to approximately 15\,km$^{2}$.
Due to projection effects the illuminated area on the ground is much larger; a simple projection with a factor of $\sec(82.8^{\circ})$ yields an illuminated area of approximately 120\,km$^{2}$.

\begin{figure}[tbp]
  \centering
  \includegraphics[width=\textwidth]{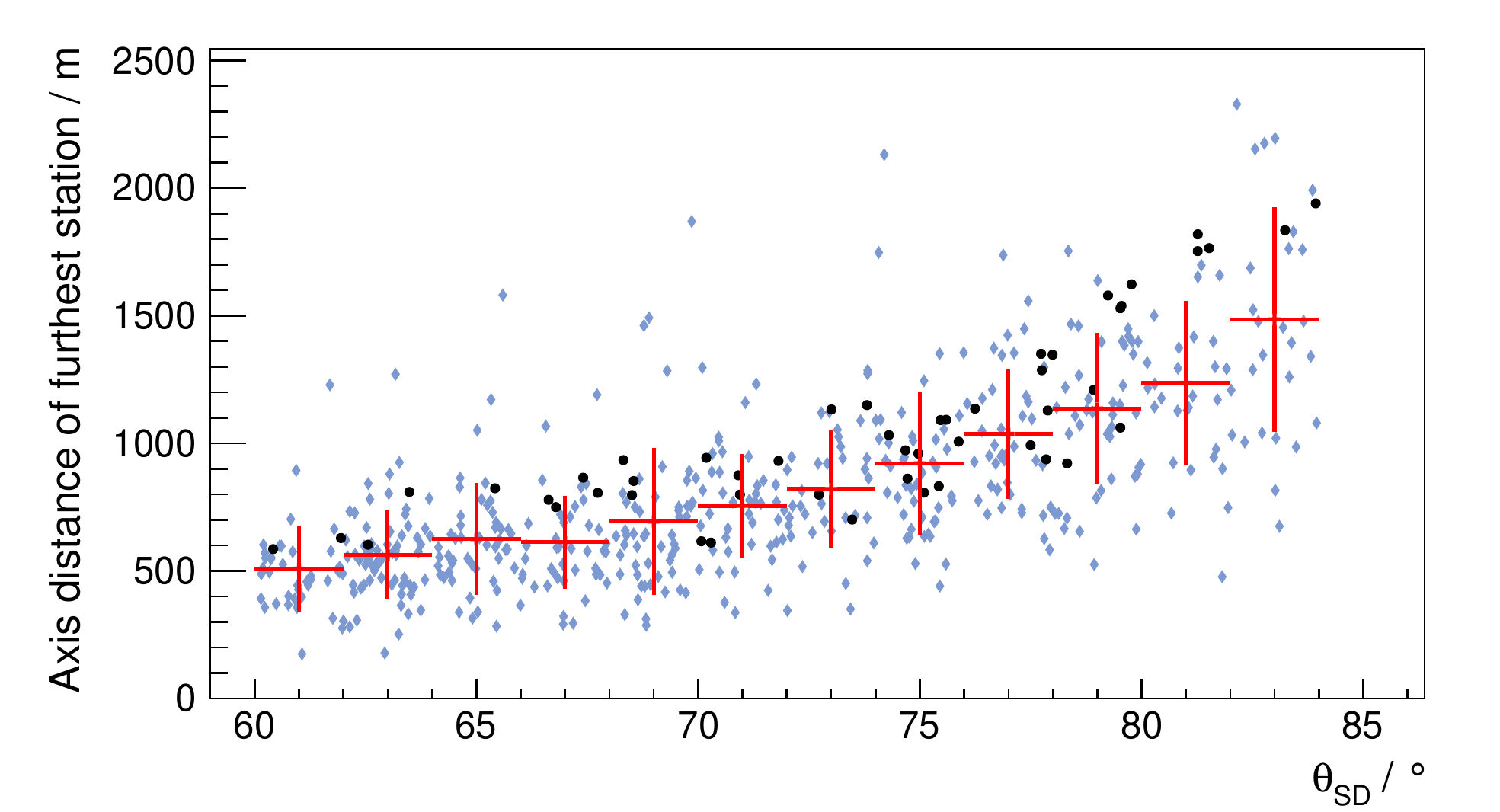}
  \caption{Farthest axis distance at which a radio signal above noise background has been detected as a function of the air-shower zenith angle. Black dots represent the 50 events that pass the quality cuts for energy reconstruction, blue diamonds denote the remaining 511 events. The red bars show the profile of the distribution, i.e., the mean and standard deviation in each $2^{\circ}$~bin. Please note that, as the radio array is significantly smaller than the radio-emission footprints, the mean values might significantly underestimate the average footprint size.}
  \label{fig:furtheststation}
\end{figure}

The distribution of the impact points of the full data set of 561 air showers illustrates that indeed many events are not contained in the geometric area of AERA, cf.\ Fig.~\ref{fig:coredistribution}.
This demonstrates that the area illuminated by radio signals is typically larger than the instrumented area of 3.5\,km$^{2}$ used in this analysis.
The farthest axis distance at which a signal-to-noise ratio of at least 10 has been measured shows a clear increase with increasing zenith angle of the air shower, as is shown in Fig.~\ref{fig:furtheststation}.
This is in qualitative agreement with forward-beamed radio emission from a receding source in the absence of absorption and scattering in the atmosphere, as explained in the introduction.
It is also in line with the observed increase in the number of detected air showers as a function of $\sin^2\theta$, i.e., an increase in detection efficiency with rising zenith angle, shown in Fig.~\ref{fig:angulardistribution}.
A weak correlation of the farthest distance with the energy of the cosmic ray is also observed and can be explained by the expected increase of the detection threshold with increasing zenith angle.

\begin{figure}[htb]
  \centering
  \includegraphics[width=0.75\textwidth]{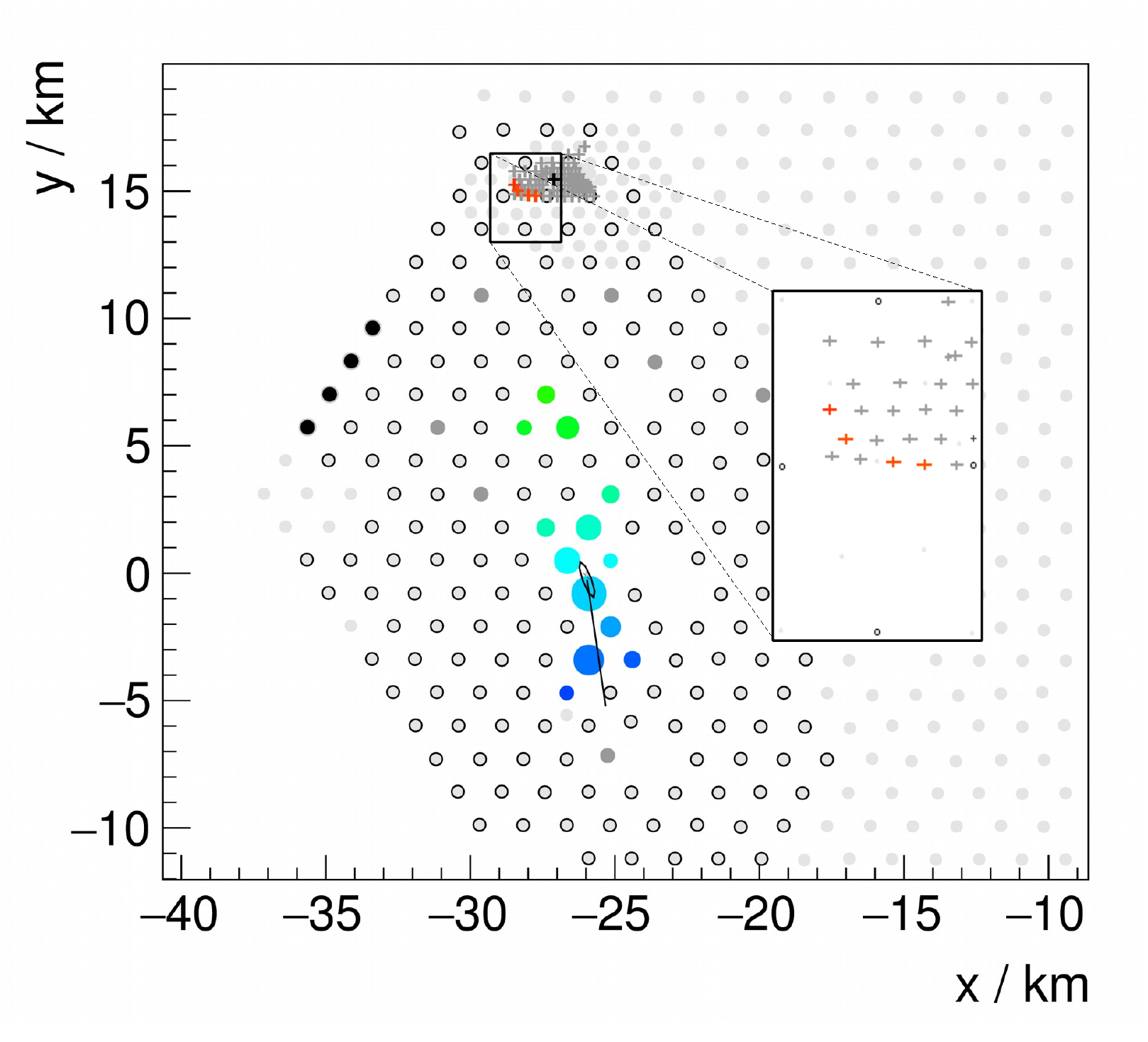}
  \caption{View of the southern-most event visible in Fig.~\ref{fig:coredistribution}. The blue to green circles indicate the measurement with the surface dector, size indicating energy deposit and color encoding arrival time. Dark-grey circles indicate isolated particle detections rejected in the reconstruction. The radio signal extends over a significantly larger area than the particle distribution.}
  \label{fig:farevent}
\end{figure}

Fig.~\ref{fig:farevent} shows a closer look at another interesting air-shower event, the southernmost one located at $(x,y) = (-26,0)$\,km in Fig.~\ref{fig:coredistribution}.
The air shower has been detected with four antennas at the edge of AERA.
Its readout was triggered because an isolated surface-detector station with significant energy deposit (dark-grey circles in Fig.~\ref{fig:farevent}) was closer than the 5\,km maximum readout distance discussed in section \ref{sec:setup}.
The locations of the antennas with a signal are in alignment with the ground projection of the air-shower axis reconstructed from the surface-detector data.
The azimuth angles reconstructed from the radio signals and particle-detector measurements agree to within better than 0.5$^{\circ}$. The zenith angle reconstructed with the particle detectors amounts to 83$^{\circ}$, while the zenith angle determined from the arrival times of the radio signals corresponds to 87$^{\circ}$.
The low number of radio antennas with signal and their approximate alignment along a line perpendicular to the air-shower axis limit the zenith-angle resolution of the radio measurement in this particular case.
It has been reported that signal reflections at the ground, which are implicitly included in our antenna models but not yet treated explicitly in our analysis, might also adversely affect the zenith-angle reconstruction at low elevations~\citep{ANITAHRA}.
The signals measured in the individual antennas have typical characteristics of air-shower radio signals (pulse shape and width as well as relative amplitude).
The maximum axis distance at which a signal has been measured amounts to 2150\,m, a value similar to that measured in other air showers; i.e., the exceptionally large ground distance arises from projection effects.
Nevertheless, this example illustrates that the ground area illuminated by radio signals can be significantly larger than the ``particle footprint'' on the ground.

\subsection{Comparison with simulations} \label{sec:simcomparison}

\begin{figure}
\centering
\begin{subfigure}[b]{.48\linewidth}
\includegraphics[width=\linewidth]{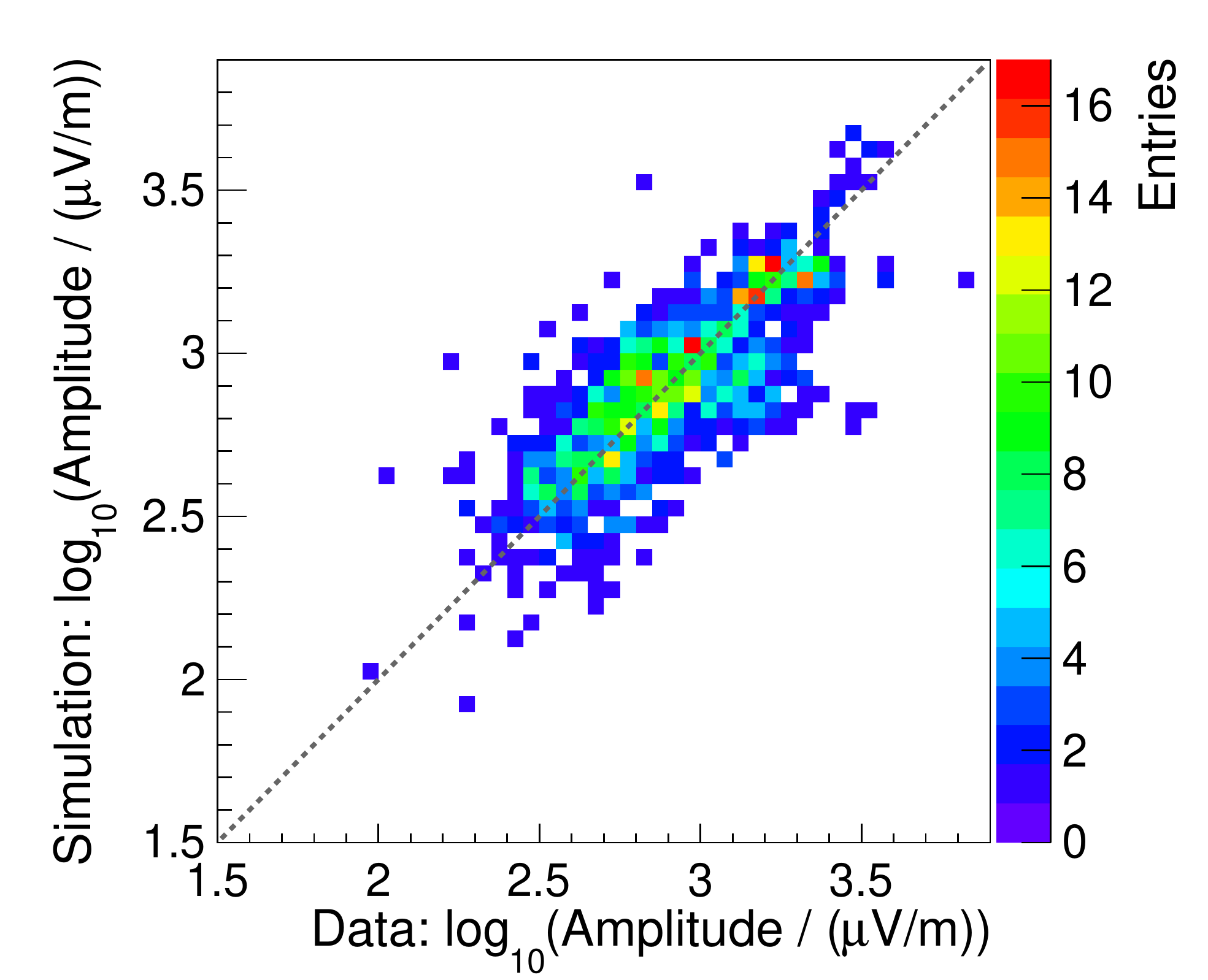}
\subcaption{Each of the 1078 entries characterizes one individual antenna station. The color scale denotes the number of entries in each bin.}
\label{fig:correlations}
\end{subfigure}%
\hfill%
\begin{subfigure}[b]{.48\linewidth}
\includegraphics[width=\linewidth]{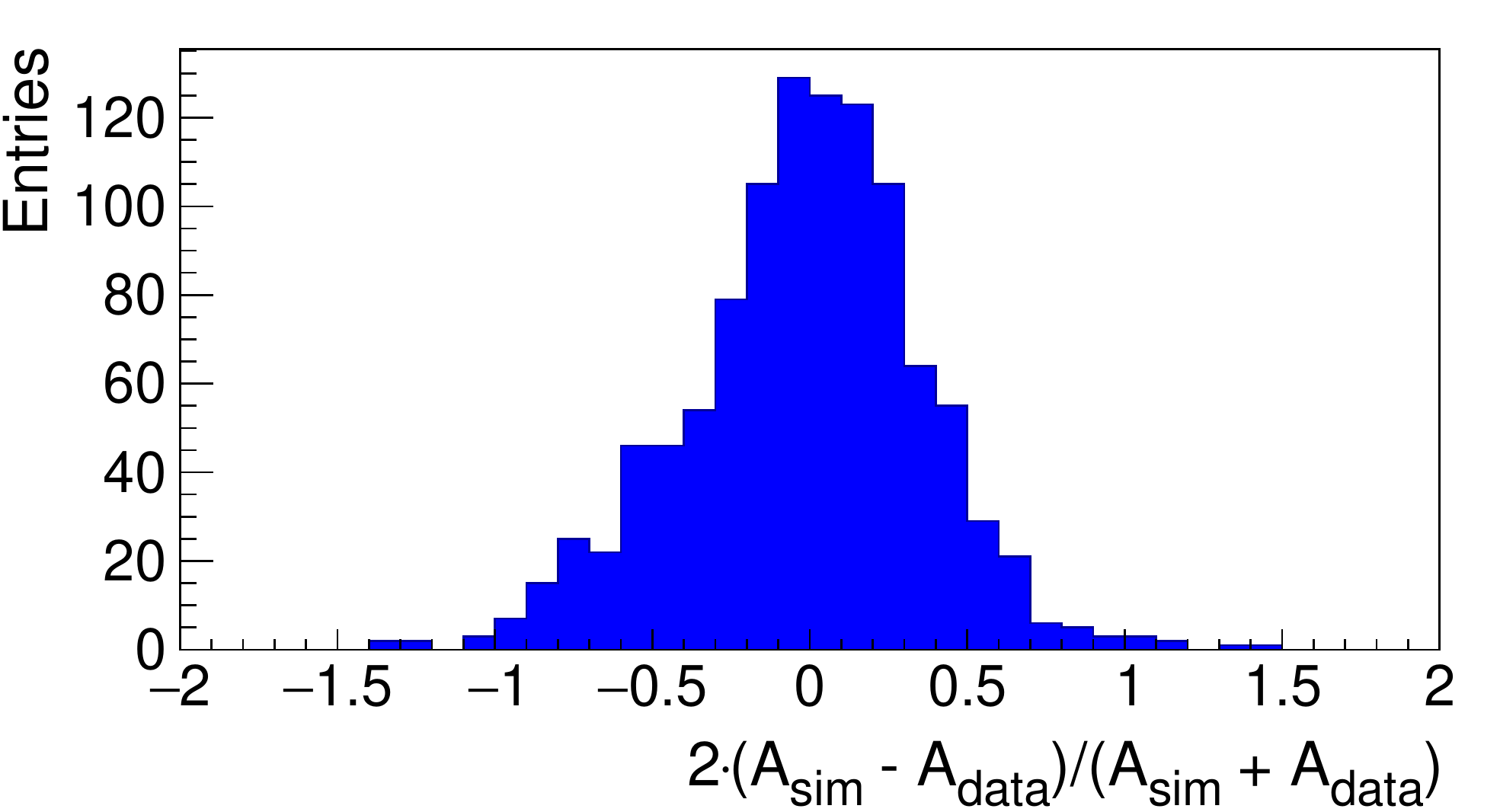}
\subcaption{Histogram of the deviations of the simulated and measured electric field amplitudes in each individual radio-detector station. The mean deviation amounts to $-2\%$, the spread as measured by the standard deviation is 38\%.}
\label{fig:deviations}
\end{subfigure}
\caption{Correlation between the CoREAS-simulated and the measured amplitudes of the electric field pulses (projected onto the horizontal plane) for the 50 measured air showers with a reconstructed energy.}
\label{fig:amp_compare}
\end{figure}

For the subset of 50 events with a surface-detector reconstruction of the cosmic-ray energy, we have made a direct comparison with the associated CoREAS-simulations.
In Fig.~\ref{fig:correlations}, we compare the simulated pulse amplitude as predicted for a given antenna station with the measured pulse amplitude in that antenna station.
Only antenna stations for which both the measured and simulated signals  pass the signal-to-noise cut of 10 are used in this comparison.
There is a clear correlation even though there is significant scatter.
Fig.~\ref{fig:deviations} shows a histogram of the corresponding relative deviation between simulated and measured amplitudes.
On average, the simulations underpredict the measured amplitudes by 2\%, which is well inside the systematic uncertainty of ${\sim}20$\% arising from the 14\% uncertainty in the energy scale of the Pierre Auger Observatory~\citep{EnergyScaleICRC2013} and the ${\sim}10$ to 15\% absolute calibration uncertainty of the two different types of AERA antennas~\citep{AERALPDA2017,AERAAntennaPaper2012}.
(We note that these antenna calibration uncertainties were determined for zenith angles up to 60$^{\circ}$~\citep{AERALPDA2017} and work is currently ongoing to quantify the uncertainties at larger zenith angles.)
The spread of 38\% is larger than observed for near-vertical air showers, however this is explainable, among other factors, by the increased uncertainty of the reconstructed impact point of inclined air showers, which is important input to the simulations.
There is thus still room for improvement when employing a detailed reconstruction of the radio signals of inclined air showers, which is currently under investigation.

\subsection{Presence of Cherenkov signature}

\begin{figure}[tbp]
  \centering
  \includegraphics[width=0.75\textwidth]{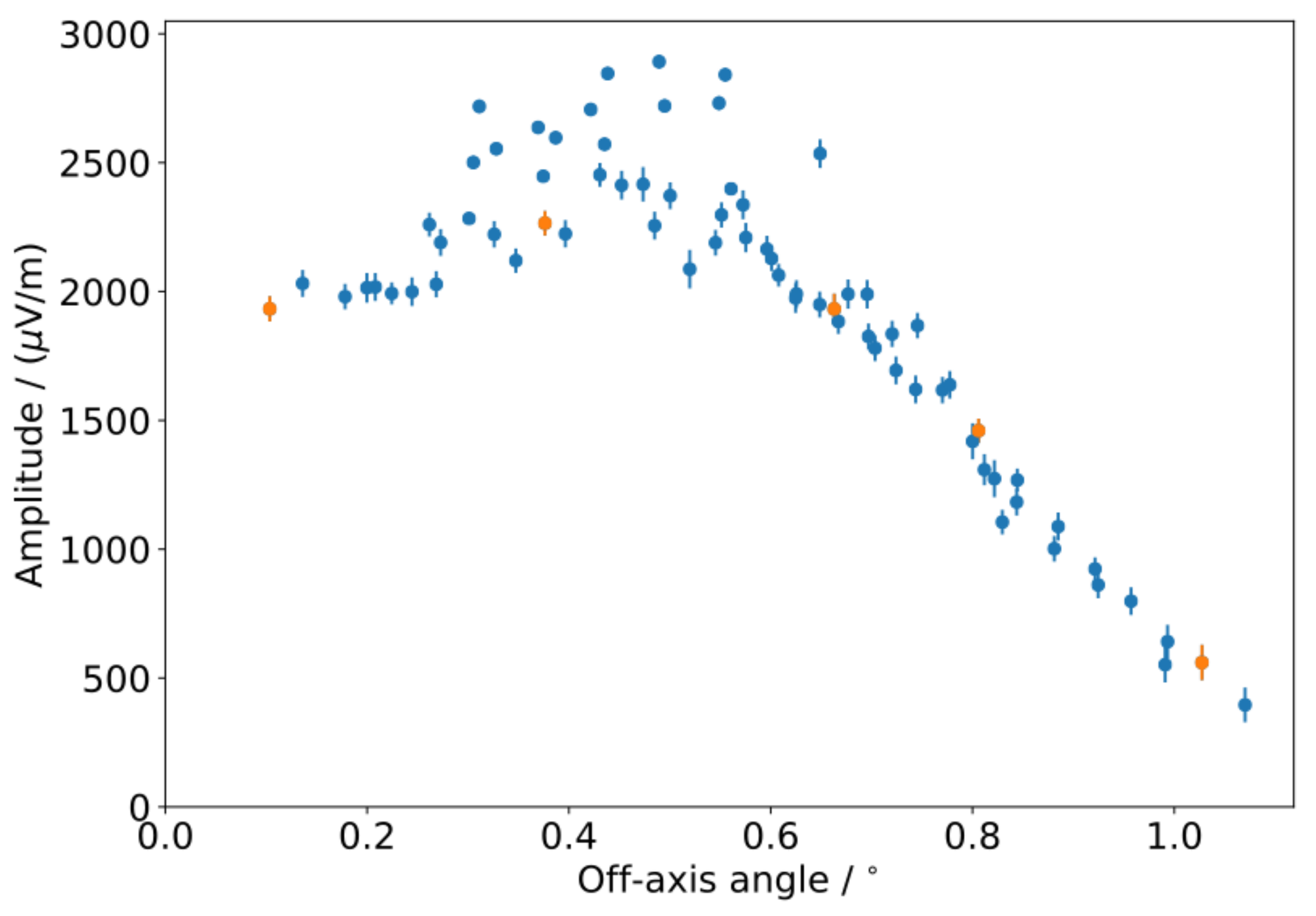}
  \caption{Distribution of amplitudes measured in the example event shown in Fig.~\ref{fig:exampleevent} (right) versus the off-axis angle for an assumed point source at an atmospheric depth of 780\,g/cm$^2$. The maximum is located at an approximate angle of 0.5$^\circ$. The orange points indicate the sampling of the event with a sparse, thinned-out AERA array (see section~\ref{sec:sparse}).
  }
  \label{fig:offaxisangle}
\end{figure}

The example event shown in Fig.~\ref{fig:exampleevent} exhibits a clear maximum in the lateral signal distribution at an axis distance of approximately 1000\,m.
Assuming that the emission arises predominantly from the shower maximum, we can estimate the off-axis angle under which this feature is seen.
At the event energy of $2{\times}10^{19}$\,eV, the average depth of shower maximum measured with the Auger fluorescence detector~\citep{AugerXmaxMoments2014,Bellido:2017cgf} amounts to ${\sim}780$\,g/cm$^2$.
Using an average density profile for the atmosphere above the observatory~\citep{Abreu:2012zg}, we relate this depth of shower maximum $X_{\mathrm{max}}$ to a geometrical source distance $d$ by solving the equation
\begin{equation}
X_0 - X_{\mathrm{max}} = \int_0^d \rho(l)\,\mathrm{d}l
\end{equation}
for $d$. Here, $X_0$ denotes the atmospheric depth of the observatory level, and $\rho(l)$ denotes the atmospheric density at the distance $l$ measured along the shower axis from the impact point to the shower maximum.
For inclined air showers, the atmospheric curvature needs to be taken into account, therefore the above equation can in general not be solved analytically and $d$ is determined numerically.
For a depth of shower maximum of 780\,g/cm$^2$ and the event zenith angle of 82.8$^{\circ}$, the geometric source distance $d$ amounts to 116\,km.
From this geometrical distance and the axis distances of the antennas, an off-axis angle for each antenna is then calculated using trigonometric relations, the result of which is shown in Fig.~\ref{fig:offaxisangle}.
The maximum in the lateral signal distribution corresponds to an off-axis angle of ${\sim}0.5^\circ$.
This value is in agreement with the angular scale on which a Cherenkov ring is expected for inclined air showers~\citep{AlvarezMunizANITASims}. The ring arises from the relativistic time-compression of the radio pulses due to the non-unity refractive index of the atmosphere~\citep{HuegePLREP}.

The derivation presented here is only very approximate and intended to illustrate the general principle.
For the majority of the events in our analysis, the Cherenkov signature is washed out by the uncertainty of the position of the air-shower impact point and by signal asymmetries arising from polarization and geometric effects.
For a reliable recovery of the Cherenkov signature, a detailed reconstruction of inclined air showers from their radio signals will thus need to be performed, which is currently under investigation.

We note that the opening angle of the Cherenkov ring is related to the depth of shower maximum~\citep{AllanCherenkov1971,DeVriesBergScholten2011,AlvarezMunizCarvalhoZas2012}.
For inclined air showers with geometrical source distances of order 100\,km, the relative effects of a varying depth of shower maximum on the signal distribution will, however, be notedly smaller than for near-vertical geometries:
For the example event presented here, a shift of $\pm$30\,g/cm$^2$ around the value of 780\,g/cm$^2$ amounts to only $\sim 2$\% variation in the geometrical source distance, whereas for an equivalent shower with 30\,$^\circ$ zenith angle the geometrical source distance varies by $\sim 10$\%.

\section{Prospects} \label{sec:sparse}

Measurements of inclined air showers using radio arrays have significant potential.

First, antenna grid spacings that are of the same order as those for particle detector arrays aiming at the measurement of the highest-energy cosmic rays should suffice for the radio detection of inclined air showers.
Even a 1.5\,km grid of radio antennas, tailored to the detection of inclined air showers, seems feasible with radio-emission footprints covering dozens of km$^{2}$ on the ground.
We illustrate this by thinning out the grid of AERA antennas used for the radio analysis to a mere five antennas on an approximate 1.5\,km grid and re-running the complete analysis procedure on the full set of raw data.

\begin{figure}[tbp]
  \centering
  \includegraphics[width=0.8\textwidth]{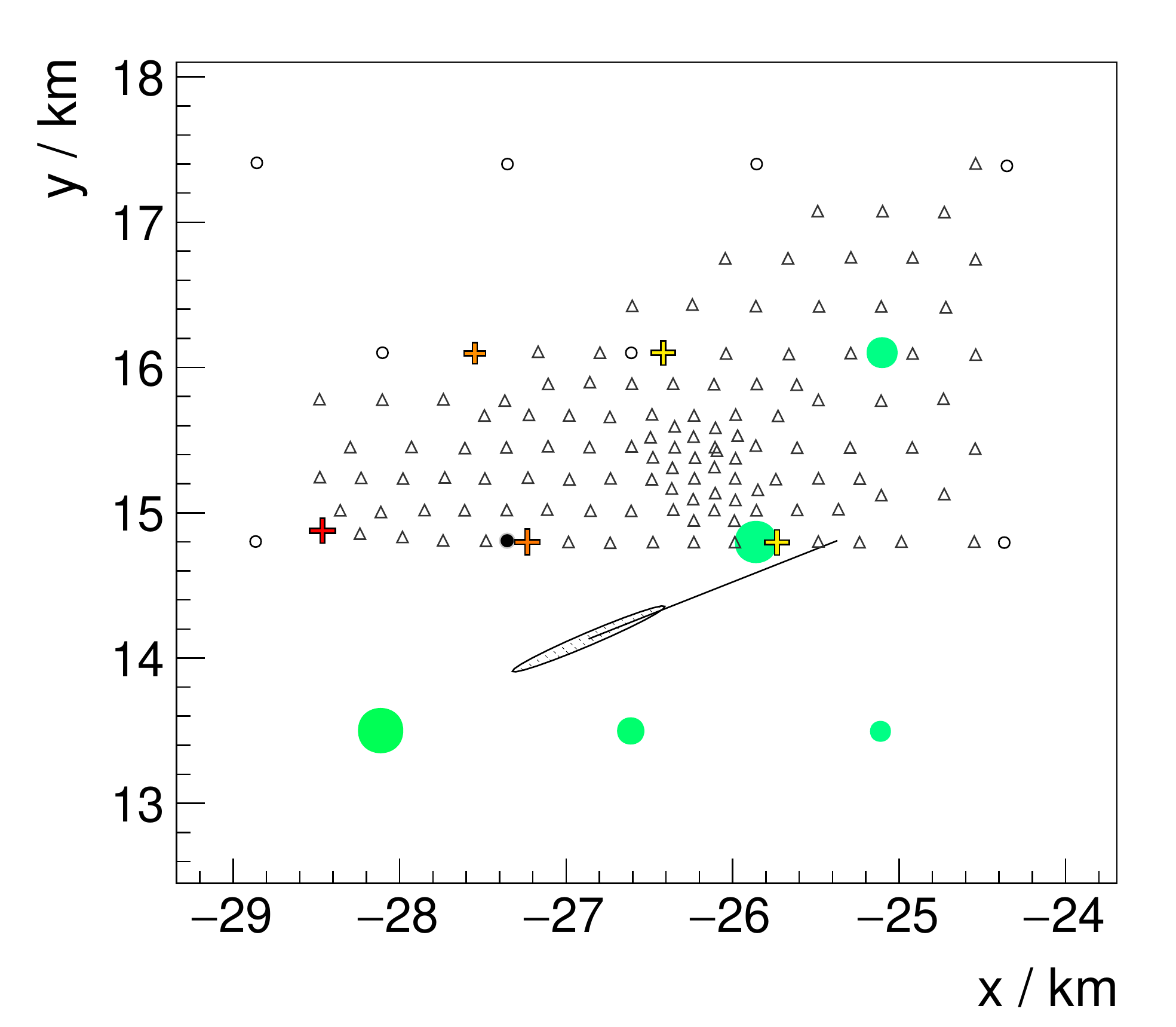}
  \caption{
  The same air shower as shown in Fig.~\ref{fig:exampleevent} but analysed by only using signals from five AERA antennas on an approximate 1.5\,km grid. All five antenna stations have a signal above threshold as marked by the colored crosses.
  }
  \label{fig:sparsereconstruction}
\end{figure}

This results in 44 air showers fulfilling the same criteria as the originally selected 561 events, including the example event previously shown in Fig.~\ref{fig:exampleevent}.
The detection of this event with a sparse array is illustrated in Fig.~\ref{fig:sparsereconstruction}.
Even with this small and sparse array a reasonable sampling of the lateral signal distribution is achieved, as is illustrated by the orange points in Fig.~\ref{fig:offaxisangle}.
For the complete set of 44 events, the reconstruction of the surface-detector data yields a mean zenith angle of $78^{\circ}$.
This is larger than the mean zenith angle of $71^{\circ}$ obtained for the 561 events observed with the full AERA as the detection probability for the sparse array increases with zenith angle.
The arrival directions reconstructed with the surface-detector and radio-detector data agree on average within $1.8^{\circ}$, which constitutes only a minor loss in accuracy compared to the full array.

Instrumentation of a radio-detector array on a 1.5\,km grid would allow direct integration of hardware in the existing surface detector array of the Pierre Auger Observatory.
Such an approach would profit from re-using much of the infrastructure (photovoltaic systems, communications hardware, local trigger) and thus dramatically reduce the cost for deployment of a large-scale radio-detector array.

Second, radio detection has the advantage of directly measuring the energy in the electromagnetic cascade of an extensive air shower~\citep{AERAEnergyPRL}.
As the radio signal is not absorbed or scattered significantly in the atmosphere, this is true independent of zenith angle~\citep{GlaserRadEnergyStudy} and, thus, also for inclined air showers.
Arrays of volumetric particle detectors, in contrast, perform an almost pure measurement of the muonic component of inclined air showers, as the electromagnetic shower has died out when the shower reaches the ground.
A combination of radio and particle detectors measuring inclined air showers thus offers significant potential for mass-composition measurements~\citep{HoltICRC2017} and studies of air-shower physics such as the currently unsatisfactory reproduction of the muonic component of extensive air showers in state-of-the-art hadronic interaction models~\citep{AugerMuonExcess}.

Finally, the detection of a large radio-emission footprint signifies that the radio-emission source is far away and the shower is ``old'' when reaching the ground.
Neutrino searches with inclined air showers detected by the Auger surface detector~\citep{Aab:2015kma} rely on the unambiguous classification of ``old'' (hadronic) versus ``young'' (neutrino-induced) inclined air showers.
Radio detection could thus provide valuable additional information to such neutrino searches:
If an air shower does exhibit a large radio-emission footprint, it can be excluded as a ``young'' air shower.
Detection of near-horizontal air showers arising from neutrino interactions in mountain ridges, as proposed in the context of the GRAND project~\citep{Fang:2017mhl}, also holds significant potential.\enlargethispage{\baselineskip}

\section{Conclusion}

We have collected a significant data set of inclined air showers at EeV energies using the Auger Engineering Radio Array and the Auger surface detector.
These showers illuminate areas of several km$^{2}$ on the ground with measurable radio signals, and the size of the emission footprint clearly increases with shower zenith angle, as expected for forward-beamed emission which does not suffer from absorption or scattering in the atmosphere.
This confirms long-standing predictions that inclined air showers should be particularly favorable for radio detection.
Per-event CoREAS-simulations of the electric-field amplitudes in individual antennas are in agreement with the measurements.
The presence of a Cherenkov ring illustrates the potential for the determination of the depth of shower maximum.
It is clear though that further work will be needed to improve the experimental accuracy.
In particular, the absolute calibration of the used radio antennas for elevations below 30$^\circ$ needs to be studied in detail.
Also, a good understanding of the lateral radio-signal distribution of inclined air showers will be needed for reliable determination of cosmic-ray parameters from radio measurements of inclined air showers.

The potential of these measurements lies in the possibility of a cost-effective instrumentation of the large areas needed for measurements at the highest cosmic-ray energies with radio antennas as well as in the complementarity of the radio and particle-detector measurements of inclined air showers.

This concept will be explored further with the latest stage of AERA, in which an additional 29 detector stations have been deployed, mostly on a grid with 750\,m distance, thereby instrumenting a total area of 17\,km$^{2}$ with 153 antenna stations~\citep{SchulzIcrc2015}. Furthermore, we envisage the direct integration of radio antennas with the Auger surface detector stations over a large part of
the observatory.

% created on 2018-06-18

\section*{Acknowledgments}

\begin{sloppypar}
The successful installation, commissioning, and operation of the Pierre
Auger Observatory would not have been possible without the strong
commitment and effort from the technical and administrative staff in
Malarg\"ue. We are very grateful to the following agencies and
organizations for financial support:
\end{sloppypar}

\begin{sloppypar}
Argentina -- Comisi\'on Nacional de Energ\'\i{}a At\'omica; Agencia Nacional de
Promoci\'on Cient\'\i{}fica y Tecnol\'ogica (ANPCyT); Consejo Nacional de
Investigaciones Cient\'\i{}ficas y T\'ecnicas (CONICET); Gobierno de la
Provincia de Mendoza; Municipalidad de Malarg\"ue; NDM Holdings and Valle
Las Le\~nas; in gratitude for their continuing cooperation over land
access; Australia -- the Australian Research Council; Brazil -- Conselho
Nacional de Desenvolvimento Cient\'\i{}fico e Tecnol\'ogico (CNPq);
Financiadora de Estudos e Projetos (FINEP); Funda\c{c}\~ao de Amparo \`a
Pesquisa do Estado de Rio de Janeiro (FAPERJ); S\~ao Paulo Research
Foundation (FAPESP) Grants No.~2010/07359-6 and No.~1999/05404-3;
Minist\'erio da Ci\^encia, Tecnologia, Inova\c{c}\~oes e Comunica\c{c}\~oes (MCTIC);
Czech Republic -- Grant No.~MSMT CR LG15014, LO1305, LM2015038 and
CZ.02.1.01/0.0/0.0/16\_013/0001402; France -- Centre de Calcul
IN2P3/CNRS; Centre National de la Recherche Scientifique (CNRS); Conseil
R\'egional Ile-de-France; D\'epartement Physique Nucl\'eaire et Corpusculaire
(PNC-IN2P3/CNRS); D\'epartement Sciences de l'Univers (SDU-INSU/CNRS);
Institut Lagrange de Paris (ILP) Grant No.~LABEX ANR-10-LABX-63 within
the Investissements d'Avenir Programme Grant No.~ANR-11-IDEX-0004-02;
Germany -- Bundesministerium f\"ur Bildung und Forschung (BMBF); Deutsche
Forschungsgemeinschaft (DFG); Finanzministerium Baden-W\"urttemberg;
Helmholtz Alliance for Astroparticle Physics (HAP);
Helmholtz-Gemeinschaft Deutscher Forschungszentren (HGF); Ministerium
f\"ur Innovation, Wissenschaft und Forschung des Landes
Nordrhein-Westfalen; Ministerium f\"ur Wissenschaft, Forschung und Kunst
des Landes Baden-W\"urttemberg; Italy -- Istituto Nazionale di Fisica
Nucleare (INFN); Istituto Nazionale di Astrofisica (INAF); Ministero
dell'Istruzione, dell'Universit\'a e della Ricerca (MIUR); CETEMPS Center
of Excellence; Ministero degli Affari Esteri (MAE); Mexico -- Consejo
Nacional de Ciencia y Tecnolog\'\i{}a (CONACYT) No.~167733; Universidad
Nacional Aut\'onoma de M\'exico (UNAM); PAPIIT DGAPA-UNAM; The Netherlands
-- Ministry of Education, Culture and Science; Netherlands Organisation
for Scientific Research (NWO); Dutch national e-infrastructure with the
support of SURF Cooperative; Poland -- National Centre for Research and
Development, Grant No.~ERA-NET-ASPERA/02/11; National Science Centre,
Grants No.~2013/08/M/ST9/00322, No.~2016/23/B/ST9/01635 and No.~HARMONIA
5--2013/10/M/ST9/00062, UMO-2016/22/M/ST9/00198; Portugal -- Portuguese
national funds and FEDER funds within Programa Operacional Factores de
Competitividade through Funda\c{c}\~ao para a Ci\^encia e a Tecnologia
(COMPETE); Romania -- Romanian Ministry of Research and Innovation
CNCS/CCCDI-UESFISCDI, projects
PN-III-P1-1.2-PCCDI-2017-0839/19PCCDI/2018, PN-III-P2-2.1-PED-2016-1922,
PN-III-P2-2.1-PED-2016-1659 and PN18090102 within PNCDI III; Slovenia --
Slovenian Research Agency; Spain -- Comunidad de Madrid; Fondo Europeo
de Desarrollo Regional (FEDER) funds; Ministerio de Econom\'\i{}a y
Competitividad; Xunta de Galicia; European Community 7th Framework
Program Grant No.~FP7-PEOPLE-2012-IEF-328826; USA -- Department of
Energy, Contracts No.~DE-AC02-07CH11359, No.~DE-FR02-04ER41300,
No.~DE-FG02-99ER41107 and No.~DE-SC0011689; National Science Foundation,
Grant No.~0450696; The Grainger Foundation; Marie Curie-IRSES/EPLANET;
European Particle Physics Latin American Network; European Union 7th
Framework Program, Grant No.~PIRSES-2009-GA-246806; and UNESCO.
\end{sloppypar}

%\bibliographystyle{elsarticle-num}
%\bibliography{references}

\begin{thebibliography}{10}
\expandafter\ifx\csname url\endcsname\relax
  \def\url#1{\texttt{#1}}\fi
\expandafter\ifx\csname urlprefix\endcsname\relax\def\urlprefix{URL }\fi
\expandafter\ifx\csname href\endcsname\relax
  \def\href#1#2{#2} \def\path#1{#1}\fi

\bibitem{HuegePLREP}
T.~Huege,
  \href{http://dx.doi.org/http://dx.doi.org/10.1016/j.physrep.2016.02.001}{{Radio
  detection of cosmic ray air showers in the digital era}}, Physics Reports 620
  (2016) 1 -- 52.

\bibitem{SchroederReview}
F.~G. Schr\"oder,
  \href{http://dx.doi.org/http://dx.doi.org/10.1016/j.ppnp.2016.12.002}{Radio
  detection of cosmic-ray air showers and high-energy neutrinos}, Progress in
  Particle and Nuclear Physics 93 (2017) 1 -- 68.

\bibitem{GoussetRavelRoy2004}
T.~{Gousset}, O.~{Ravel}, C.~{Roy},
  \href{https://doi.org/10.1016/j.astropartphys.2004.05.004}{{Are
  vertical cosmic rays the most suitable to radio detection?}}, Astropart.
  Phys. 22 (2004) 103--107.

\bibitem{HuegeFalcke2005b}
T.~{Huege}, H.~{Falcke}, \href{http://dx.doi.org/10.1016/j.astropartphys.2005.05.008}{Radio emission from cosmic ray air showers:
  Simulation results and parametrization}, Astropart. Phys. 24 (2005) 116.

\bibitem{HuegeUHECR2014}
T.~{Huege}, A.~{Haungs}, \href{http://dx.doi.org/10.7566/JPSCP.9.010018} {Radio detection of cosmic rays: present and future},
  JPS Conference Proceedings 09 (2016) 010018, proceedings of the UHECR2014 conference, Springdale, USA.

\bibitem{PetrovicApelAsch2006}
J.~{Petrovic}, W.~D. {Apel}, T.~{Asch}, {et al.}, \href
  {http://dx.doi.org/10.1051/0004-6361:20065732}{Radio emission of highly
  inclined cosmic ray air showers measured with LOPES}, Astronomy \&
  Astrophysics 462 (2007) 389--395.

\bibitem{HooverNamGorham2010}
S.~{Hoover}, J.~{Nam}, P.~W. {Gorham}, {et al.},
  \href{http://dx.doi.org/10.1103/PhysRevLett.105.151101}{{Observation of
  Ultrahigh-Energy Cosmic Rays with the ANITA Balloon-Borne Radio
  Interferometer}}, Phys. Rev. Lett. 105 (2010) 151101.

\bibitem{ANITAEnergy}
H.~Schoorlemmer, et~al., \href{http://dx.doi.org/10.1016/j.astropartphys.2016.01.001}{Energy and Flux Measurements of Ultra-High Energy
  Cosmic Rays Observed During the First ANITA Flight}, Astropart. Phys. 77
  (2016) 32--43.

\bibitem{Barwick:2016mxm}
S.~W. Barwick, et~al., \href{http://dx.doi.org/10.1016/j.astropartphys.2017.02.003} {Radio detection of air showers with the ARIANNA
  experiment on the Ross Ice Shelf}, Astropart. Phys. 90 (2017) 50--68.

\bibitem{SchulzIcrc2015}
J.~{Schulz} {for the Pierre Auger Collaboration}, \href{https://doi.org/10.22323/1.236.0615} {Status and Prospects of the
  Auger Engineering Radio Array}, in: Proceedings of the 34th ICRC, The Hague,
  The Netherlands, no. PoS(ICRC2015)615, 2015.

\bibitem{AugerNIM2014}
{Pierre Auger collaboration}, A.~{Aab} {et al.}, \href{https://doi.org/10.1016/j.nima.2015.06.058} {The Pierre Auger Cosmic Ray
  Observatory}, Nucl. Instrum. Meth. A 798 (2015) 172--213.

\bibitem{AERAAntennaPaper2012}
{Pierre Auger collaboration}, P.~{Abreu} {et al.}, \href{http://dx.doi.org/10.1088/1748-0221/7/10/P10011} {Antennas for the detection
  of radio emission pulses from cosmic-ray induced air showers at the Pierre
  Auger Observatory}, JINST 7 (2012) P10011.

\bibitem{AERALPDA2017}
{Pierre Auger Collaboration}, A.~{Aab} {et~al.}, \href
  {http://10.1088/1748-0221/12/10/T10005} {Calibration of the
  Logarithmic-Periodic Dipole Antenna (LPDA) Radio Stations at the Pierre Auger
  Observatory using an Octocopter}, JINST 12 (2017) T10005.

\bibitem{ApelArteagaBaehren2012a}
W.~D. {Apel}, J.~C. {Arteaga}, L.~{B\"ahren}, {et al.}, \href{https://doi.org/10.1016/j.nima.2012.08.082} {LOPES-3D, an antenna
  array for full signal detection of air-shower radio emission}, Nucl. Instr.
  Meth. A 696 (2012) 100--109.

\bibitem{ArgiroOffline2007}
S.~{Argir\'o}, S.~L.~C. {Barroso}, J.~{Gonzalez}, {et al.}, \href {http://dx.doi.org/10.1016/j.nima.2007.07.010} {The offline
  software framework of the Pierre Auger Observatory}, Nucl. Instr. and Meth. A
  580 (2007) 1485--1496.

\bibitem{AugerSDHAS}
{Pierre Aguer collaboration}, A.~{Aab} {et al.}
  \href{https://doi.org/10.1088/1475-7516/2014/08/019}{{Reconstruction of
  inclined air showers detected with the Pierre Auger Observatory}}, Journal of
  Cosmology and Astroparticle Physics 2014~(08) (2014) 019.

\bibitem{KambeitzARENA2016}
O.~{Kambeitz},
  \href{https://doi.org/10.1051/epjconf/201713501015}{Measurement of horizontal
  air showers with the Auger Engineering Radio Array}, EPJ Web Conf. 135 (2017)
  01015.

\bibitem{KambeitzThesis2016}
O.~{Kambeitz}, \href{http://nbn-resolving.org/urn:nbn:de:swb:90-557582}{{Radio
  Detection of Horizontal Extensive Air Showers}}, Ph.D. thesis, Karlsruhe
  Institute of Technology (2016).

\bibitem{Abraham:2010zz}
{Pierre Auger collaboration}, J.~Abraham {et~al.}, \href{http://dx.doi.org/10.1016/j.nima.2009.11.018} {Trigger and aperture of the surface detector array of the
  Pierre Auger Observatory}, Nucl. Instrum. Meth. A613 (2010) 29--39.

\bibitem{DembinskiThesis2009}
H.~{Dembinski}, \href{https://web.physik.rwth-aachen.de/~hebbeker/theses/dembinski_phd.pdf} {Measurement of the flux of ultra high energy cosmic rays using
  data from very inclined air showers at the Pierre Auger Observatory}, Ph.D.
  thesis, RWTH Aachen University (2009).

\bibitem{HuegeARENA2012a}
T.~{Huege}, M.~{Ludwig}, C.~W. {James}, \href{https://doi.org/10.1063/1.4807534} {Simulating radio emission from air
  showers with CoREAS}, AIP Conf. Proc.~(1535) (2013) 128--132.

\bibitem{HeckKnappCapdevielle1998}
D.~{Heck}, J.~{Knapp}, J.~N. {Capdevielle}, G.~{Schatz}, T.~{Thouw}, \href{http://digbib.ubka.uni-karlsruhe.de/volltexte/fzk/6019/6019.pdf} {CORSIKA:
  A Monte Carlo Code to Simulate Extensive Air Showers}, FZKA Report 6019,
  Forschungszentrum Karlsruhe (1998).

\bibitem{Ostapchenko:2010vb}
S.~Ostapchenko,   \href {http://dx.doi.org/10.1103/PhysRevD.83.014018}
 {Monte Carlo treatment of hadronic interactions in enhanced
  Pomeron scheme: I. QGSJET-II model}, Phys. Rev. D83 (2011) 014018.

\bibitem{Bleicher:1999xi}
M.~Bleicher, et~al., \href
  {http://dx.doi.org/10.1088/0954-3899/25/9/308}{Relativistic hadron hadron collisions in the
  ultrarelativistic quantum molecular dynamics model}, J. Phys. G25 (1999)
  1859--1896.

\bibitem{Pierog:2015epa}
T.~Pierog, \href {http://dx.doi.org/10.1051/epjconf/20159909002} {Modelling hadronic interactions in cosmic ray Monte Carlo
  generators}, EPJ Web Conf. 99 (2015) 09002.

\bibitem{AbreuAgliettaAhn2011}
{Pierre Auger collaboration}, P.~{Abreu} {et al.},
  \href {http://dx.doi.org/10.1016/j.nima.2011.01.049}{{Advanced
  functionality for radio analysis in the Offline software framework of the
  Pierre Auger Observatory}}, Nucl. Instr. Meth. A 635 (2011) 92--102.

\bibitem{AERAEnergyPRD}
{Pierre Auger collaboration}, {A.~Aab} {et al.},
  \href {http://dx.doi.org/10.1103/PhysRevD.93.122005}{{Energy estimation
  of cosmic rays with the Engineering Radio Array of the Pierre Auger
  Observatory}}, Phys. Rev. D 93 (2016) 122005.

\bibitem{ANITAHRA}
{TAROGE and ARIANNA Collaborations}, {Shih-Hao~Wang} {et al.},
  \href{http://dx.doi.org/10.22323/1.301.0358}{{Calibration, Performance,
  and Cosmic Ray Detection of ARIANNA-HCR Prototype Station}}, in:
  Proceedings of the 35th ICRC, Busan, Korea, no. PoS ICRC2017 (2017) 358.

\bibitem{EnergyScaleICRC2013}
V.~{Verzi} {for the Pierre Auger Collaboration},
  \href{https://arxiv.org/abs/1307.5059}{{The energy scale of
  the Pierre Auger Observatory}}, Proc. 33rd ICRC, Rio de Janeiro, Brazil.

\bibitem{AugerXmaxMoments2014}
{Pierre Auger collaboration}, {A.~Aab} {et~al.},
  \href {http://dx.doi.org/10.1103/PhysRevD.90.122005}{Depth of maximum
  of air-shower profiles at the Pierre Auger observatory. i. Measurements at
  energies above $10^{17.8}$~ev}, Phys. Rev. D 90 (2014) 122005.

\bibitem{Bellido:2017cgf}
J.~Bellido {for the Pierre Auger collaboration},
  \href{http://inspirehep.net/record/1618417/files/1617990_40-47.pdf}{{Depth of
  maximum of air-shower profiles at the Pierre Auger Observatory: Measurements
  above $10^{17.2}$ eV and Composition Implications}}, in: {The Pierre Auger
  Observatory: Contributions to the 35th International Cosmic Ray Conference
  (ICRC 2017)}, 2017, pp. 40--47.

\bibitem{Abreu:2012zg}
{Pierre Auger collaboration}, P.~Abreu {et~al.},   \href{http://dx.doi.org/10.1016/j.astropartphys.2011.12.002} {Description of Atmospheric Conditions at the Pierre Auger
  Observatory using the Global Data Assimilation System (GDAS)}, Astropart.
  Phys. 35 (2012) 591--607.

\bibitem{AlvarezMunizANITASims}
J.~{Alvarez-Mu\~niz}, W.~R. {Carvalho}, A.~{Romero-Wolf}, {et al.},
  \href{http://dx.doi.org/10.1103/PhysRevD.86.123007}{{Coherent radiation
  from extensive air showers in the ultrahigh frequency band}}, Phys. Rev. D 86
  (2012) 123007.

\bibitem{AllanCherenkov1971}
H.~R. {Allan}, {The Lateral Distribution of the Radio Emission, and its
  Dependence on the Longitudinal Structure of the Air Shower.}, International
  Cosmic Ray Conference 3 (1971) 1108.

\bibitem{DeVriesBergScholten2011}
K.~D. {de Vries}, A.~M. {van den Berg}, O.~{Scholten}, K.~{Werner}, \href{http://dx.doi.org/10.1103/PhysRevLett.107.061101} {Coherent
  Cherenkov Radiation from Cosmic-Ray-Induced Air Showers}, Phys. Rev. Lett.
  107~(6) (2011) 61101.

\bibitem{AlvarezMunizCarvalhoZas2012}
J.~{Alvarez-Mu\~niz}, W.~R. {Carvalho Jr.}, E.~{Zas},
  \href {http://dx.doi.org/10.1016/j.astropartphys.2011.10.005}{{Monte
  Carlo simulations of radio pulses in atmospheric showers using ZHAireS}},
  Astropart. Phys. 35~(6) (2012) 325 -- 341.

\bibitem{AERAEnergyPRL}
A.~Aab, P.~Abreu, M.~Aglietta, {et al.},\href {http://dx.doi.org/10.1103/PhysRevLett.116.241101}{{Measurement of
  the Radiation Energy in the Radio Signal of Extensive Air Showers as a
  Universal Estimator of Cosmic-Ray Energy}}, Phys. Rev. Lett. 116 (2016)
  241101.

\bibitem{GlaserRadEnergyStudy}
C.~Glaser, M.~Erdmann, J.~R. H\"orandel, T.~Huege, J.~Schulz,
  \href{http://dx.doi.org/10.1088/1475-7516/2016/09/024}{Simulation of
  radiation energy release in air showers}, Journal of Cosmology and
  Astroparticle Physics 09 (2016) 024.

\bibitem{HoltICRC2017}
E.~M. {Holt} for {the Pierre Auger Collaboration}, \href{https://pos.sissa.it/301/492/pdf} {Recent results of the Auger
  Engineering Radio Array (AERA)}, in: Proceedings of the 35th ICRC, Busan,
  Korea, no. PoS(ICRC2017)492, 2017.

\bibitem{AugerMuonExcess}
{Pierre Auger collaboration}, A.~Aab {et al.},
  \href {http://dx.doi.org/10.1103/PhysRevD.91.032003}{{Muons in air
  showers at the Pierre Auger Observatory: Mean number in highly inclined
  events}}, Phys. Rev. D 91 (2015) 032003.

\bibitem{Aab:2015kma}
{Pierre Auger collaboration}, A.~Aab {et~al.},   \href {http://dx.doi.org/10.1103/PhysRevD.91.092008} {Improved limit to the diffuse flux of ultrahigh energy
  neutrinos from the Pierre Auger Observatory}, Phys. Rev. D91~(9) (2015)
  092008.

\bibitem{Fang:2017mhl}
K.~Fang, et~al., \href{https://pos.sissa.it/301/996} {The Giant Radio Array for Neutrino Detection (GRAND): Present
  and Perspectives}, in: Proceedings of the 35th ICRC, Busan,
  Korea, no. PoS ICRC2017 (2017) 996.

\end{thebibliography}

\end{document}